\documentclass[reprint,amsmath,aps,prl,twocolumn]{revtex4-1}
\usepackage[utf8]{inputenc}
\usepackage[T1]{fontenc}
\usepackage{graphicx}
\usepackage{dcolumn}
\usepackage{amsmath}
\usepackage{amssymb}
\usepackage{array}
\usepackage{multirow}
\usepackage{calc}
\usepackage{color}
\usepackage{graphicx}
\usepackage{float}
\usepackage{bm}
\usepackage{hyperref}

\newcommand{\icm}{\ensuremath{~\textrm{cm}^{-1}}}

\newcommand{\urusi}{URu$_2$Si$_2$}

\begin{document}

\title{Anisotropic Kondo pseudo-gap in URu$_2$Si$_2$}

\author{J. Buhot$^{1,2}$}
\email{jonathan.buhot@ru.nl}
\author{X. Montiel$^{3}$}
\author{Y. Gallais$^{2}$}
\author{M. Cazayous$^{2}$}
\author{A. Sacuto$^{2}$}
\author{G. Lapertot$^{4}$}
\author{D. Aoki$^{5,4}$}
\author{N. E. Hussey$^{1}$}
\author{C. P\'{e}pin$^{6}$}
\author{S. Burdin$^{7}$}
\author{M.-A. M\'easson$^{8,2}$}
\affiliation{$^1$High Field Magnet Laboratory (HFML-EMFL), Institute for Molecules and Materials, Radboud University Nijmegen, Toernooiveld 7, 6525 ED Nijmegen, The Netherlands\\
$^2$Laboratoire Mat\'eriaux et Ph\'enom\`enes Quantiques, UMR 7162 CNRS, Universit\'e Paris Diderot, B$\hat{a}$t. Condorcet 75205 Paris Cedex 13, France\\
$^3$Department of Physics, Royal Holloway, University of London, Egham, Surrey TW20 0EX, UK\\
$^4$Université Grenoble Alpes, CEA, INAC, PHELIQS, F-38000 Grenoble, France\\
$^5$Institute for Materials Research, Tohoku University, Oarai, Ibaraki 311-1313, Japan\\
$^6$Institut de Physique Th\'{e}orique, CEA-Saclay, 91191 Gif-sur-Yvette, France\\
$^7$Université Bordeaux, CNRS, LOMA, UMR 5798, F 33400 Talence, France\\
$^8$Institut NEEL CNRS/UGA UPR2940, MCBT, 25 rue des Martyrs BP 166, 38042 Grenoble cedex 9, France
}

\begin{abstract}

A polarized electronic Raman scattering study reveals the emergence of symmetry dependence in the electronic Raman response of single crystalline URu$_{2}$Si$_{2}$ below the Kondo crossover scale $T_K\sim$~100~K. In particular, the development of a coherent Kondo pseudo-gap predominantly in the E$_g$ channel highlights strong anisotropy in the Kondo physics in URu$_{2}$Si$_{2}$ that has previously been neglected in theoretical models of this system and more generally has been sparsely treated for Kondo systems. A calculation of the Raman vertices demonstrates that the strongest Raman vertex does indeed develop within the E$_g$ channel for interband transitions and reaches a maximum along the diagonals of the Brillouin zone, implying a \textit{d}-wave-like geometry for the Kondo pseudo-gap. Below the hidden order phase transition at $T_{HO}= 17.5$ K, the magnitude of the pseudo-gap is found to be enhanced.

\end{abstract}

\date{\today}

\maketitle

The Kondo effect, in which an antiferromagnetic coupling between conduction electrons and local quantum impurities leads to the formation of a Kondo singlet, is well understood for single Kondo impurity systems and is characterized by the appearance of a Kondo resonance in the electronic density of states below a crossover temperature $T_K$ \cite{kondo_resistance_1964}.
By contrast, Kondo lattice systems containing rare-earth elements (for which the Kondo impurities are realized by a lattice of magnetic moments stemming from the localized \textit{f} electron orbitals), appear much more complex \cite{hewson_kondo_1997} and exhibit surprising phenomena \cite{yang_emergent_2012,lonzarich_toward_2017}. Usually, the microscopic state of heavy-fermions is described with Anderson or Kondo lattice models \cite{hewson_kondo_1997} and most of the time the Kondo singlet state is considered to be fully isotropic (\textit{s}-wave symmetry with zero angular momentum). In certain circumstances, however, a momentum-dependent Anderson hybridization, for which the \textit{f} site hybridizes with its nearest-neighbor conduction sites \cite{weber_heavy-fermion_2008}, or a non-local Kondo coupling \cite{ghaemi_higher_2007} can give rise to the formation of heavy quasiparticles with a strong momentum-space anisotropy - in such a case the singlet state will have a higher angular momentum rank (\textit{p}, \textit{d}, ...). As a consequence, the Kondo pseudo-gap (partial gapping of the Fermi surface) that develops in the coherent Kondo regime \cite{burdin_multiple_2009,riseborough_mixed_2016,nozieres_kondo_2005,lacroix_phase_1979} may be strongly anisotropic, even exhibiting nodal and antinodal regions. Similarly to the widely discussed relationship between the pseudo-gap and the superconducting phase in the cuprates, an anisotropic (non \textit{s}-wave) Kondo pseudo-gap may have significant consequences for the entire phase diagram of these Kondo systems. However, reports of such momentum-dependent hybridization are scarce \cite{burch_optical_2007} and remain indirect.

A non-local Kondo hybridization may be particularly relevant for the complex Kondo system URu$_{2}$Si$_{2}$ where an enigmatic "hidden" order (HO) phase appears inside the Kondo coherent regime below $T_{HO}=17.5K$ and an unconventional superconducting (SC) phase develops below $T_{c}=1.5 K$ \cite{mydosh_colloquium:_2011, mydosh_hidden_2014-1}.
Many spectroscopic experiments have reported a Kondo pseudo-gap opening at a temperature scale $T_{K}= 50-100$ K, well above $T_{HO}$, suggesting that Kondo hybridization is not the cause of the HO transition \cite{bonn_far-infrared_1988, nagel_optical_2012,guo_hybridization_2012,levallois_hybridization_2011, park_observation_2012,bachar_detailed_2016}.
On the other hand, scanning tunneling microscopy (STM) has revealed an asymmetric pseudo-gap opening at $T_{HO}$ within the Kondo-Fano lattice structure that appears below $T_{K}$ \cite{schmidt_imaging_2010, aynajian_visualizing_2010}. The HO gap is found to be correlated on the atomic scale with the electronic signatures of the Kondo lattice state, suggesting that the two phenomena involve the same electronic states.
Angle-resolved photoemission spectroscopy (ARPES), on the other hand, has shown that a Kondo pseudo-gap opens at the $X$ point of the Brillouin zone, while the HO gap affects electronic states at the $\Gamma$ and $Z$ points \cite{boariu_momentum-resolved_2013}. Nevertheless, the electronic structure in the HO state at the $Z$ point (M-shape dispersion) results from the hybridization of a light
electron band with already hybridized bands characterizing the paramagnetic state.
Moreover, high magnetic field measurements have demonstrated that the destabilization of the HO state (around 37 T) coincides with the ''collapse'' of the Kondo crossover ($T_{\rho,max}, H_{\rho,max}\rightarrow0$) \cite{scheerer_interplay_2012}.

Significant theoretical works have been performed to identify the correct order parameter for the HO state \cite{mydosh_hidden_2014-1}.
Some theories consider the Kondo hybridization as an important ingredient in the realization of the HO \cite{sikkema_ising-kondo_1996,mineev_interplay_2005,
haule_arrested_2009,pepin_modulated_2011,riseborough_phase_2012}, with others even proposing the hybridization as the order parameter of the HO state itself \cite{dubi_hybridization_2011, chandra_hastatic_2013}. 
Moreover, under a pressure of $\sim$0.5 GPa, the SC and HO states collapse together  \cite{mydosh_colloquium:_2011}, promoting the HO as a precursor of SC pairing possibly of a chiral \textit{d}-wave type \cite{kasahara_exotic_2007,kawasaki_time-reversal_2014,akbari_hidden-order_2014-1}. The proper role played by the HO in the appearance of the SC, however, remains unclear.

In this Letter, we present Raman polarized spectroscopy measurements in both the Kondo phase and the HO phase of URu$_{2}$Si$_{2}$. We observe a Kondo pseudo-gap opening only in the E$_g$ symmetry below $T_{K}$. These results, together with our accompanying Raman calculations, highlight a strong anisotropy in the Kondo physics in \textit{k}-space and reveal a \textit{d}-wave like geometry for the Kondo pseudo-gap that may play a key role in the realization of the HO phase.

Polarized Raman experiments have been carried out using a solid-state laser emitting at 532~nm and a K$^+$ laser emitting at 647~nm. The scattered light has been analyzed by a Jobin Yvon T64000 simple grating spectrometer equipped with an ultra-steep long-pass edge filter to reject the Rayleigh scattering. Single crystals of URu$_2$Si$_2$ were grown by the Czochralski method using a tetra-arc furnace \cite{aoki_field_2010}.
All the D$_{4h}$ point group (space group n$^\circ139$) symmetries  \cite{hayes_scattering_2004} have been measured.
To probe the E$_{g}$ symmetry, we have used samples polished along the \textit{ac}-plane. For completeness, A$_{1g}$, A$_{2g}$, B$_{1g}$ and B$_{2g}$ symmetries have been explored on samples freshly cleaved along the \textit{ab}-plane. 
Temperature-dependent studies were carried out in a closed-cycle $^{4}$He cryostat with the sample in vacuum. All reported temperatures are corrected for laser heating following Refs. \cite{hackl_gap_1983, maksimov_investigations_1992, mialitsin_raman_2010, aliev_anisotropy_1991}.

\begin{figure}[h!]
\centering
\includegraphics[width=1\linewidth]{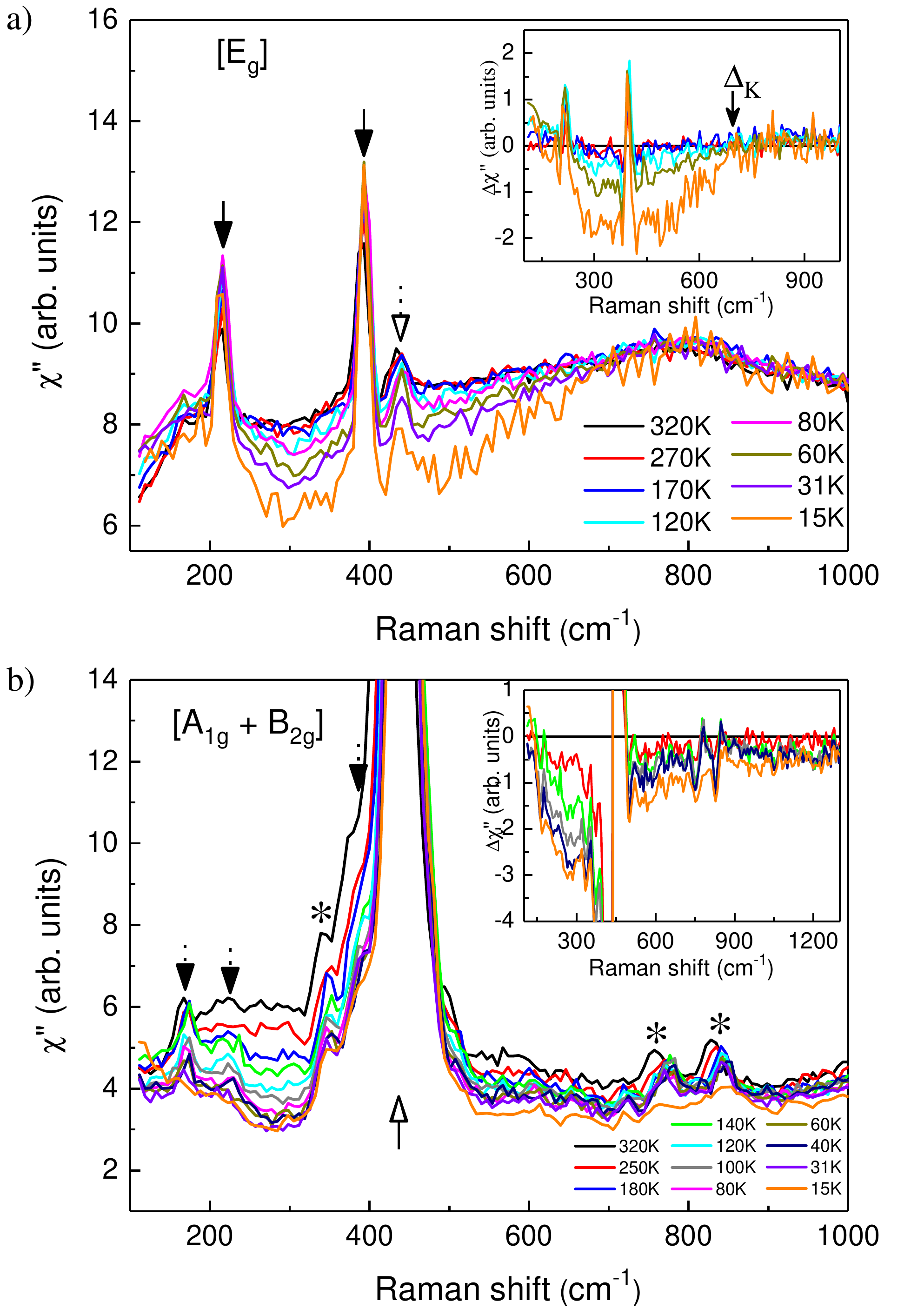}
\caption{(Color online) Raman spectra of \urusi~in the a) E$_{g}$ and b) A$_{1g}$+B$_{2g}$ symmetries. The susceptibilities $\chi''$ are normalized to the laser power. Full and open arrows indicate respectively the E$_{g}$ and A$_{1g}$ phonon modes. Dotted arrows indicate leakages of phonon modes. The symbol $\ast$ denote double phonon processes~\cite{buhot_raman_2017}. Insets show the subtracted Raman responses $\Delta\chi''=\chi''(T)-\chi''(T_{320K})$. The laser line wavelength used is 532~nm. $\Delta_{K}$ corresponds to the energy below which the Kondo pseudo-gap opens.}
\label{fig1}
\end{figure}

Figure \ref{fig1} presents the Raman spectra of \urusi~in the E$_{g}$ and A$_{1g}$+B$_{2g}$ symmetries. For the E$_{g}$ symmetry (Fig. \ref{fig1} a)), the sharp peaks at 210\icm~and 390\icm~are the two E$_{g}$ phonon modes \cite{buhot_lattice_2015}, while the small peak at 430\icm~is a leakage of the A$_{1g}$ phonon mode.
The bump at $\sim800$ \icm~has been associated with a crystal electric field excitation \cite{buhot_raman_2017}.
In the [A$_{1g}$+B$_{2g}$] symmetry (Fig. \ref{fig1} b)), the intense A$_{1g}$ phonon mode is observed at $\sim 430\icm$.
Smaller features seen at 350, 760 and 832~\icm~are related to double phonon processes \cite{buhot_raman_2017}. Finally, a tiny leakage of B$_{1g}$ and E$_{g}$ phonon modes are visible respectively at 165, 210 and 390\icm.

\begin{figure}[h!]
\centering
\includegraphics[width=1\linewidth]{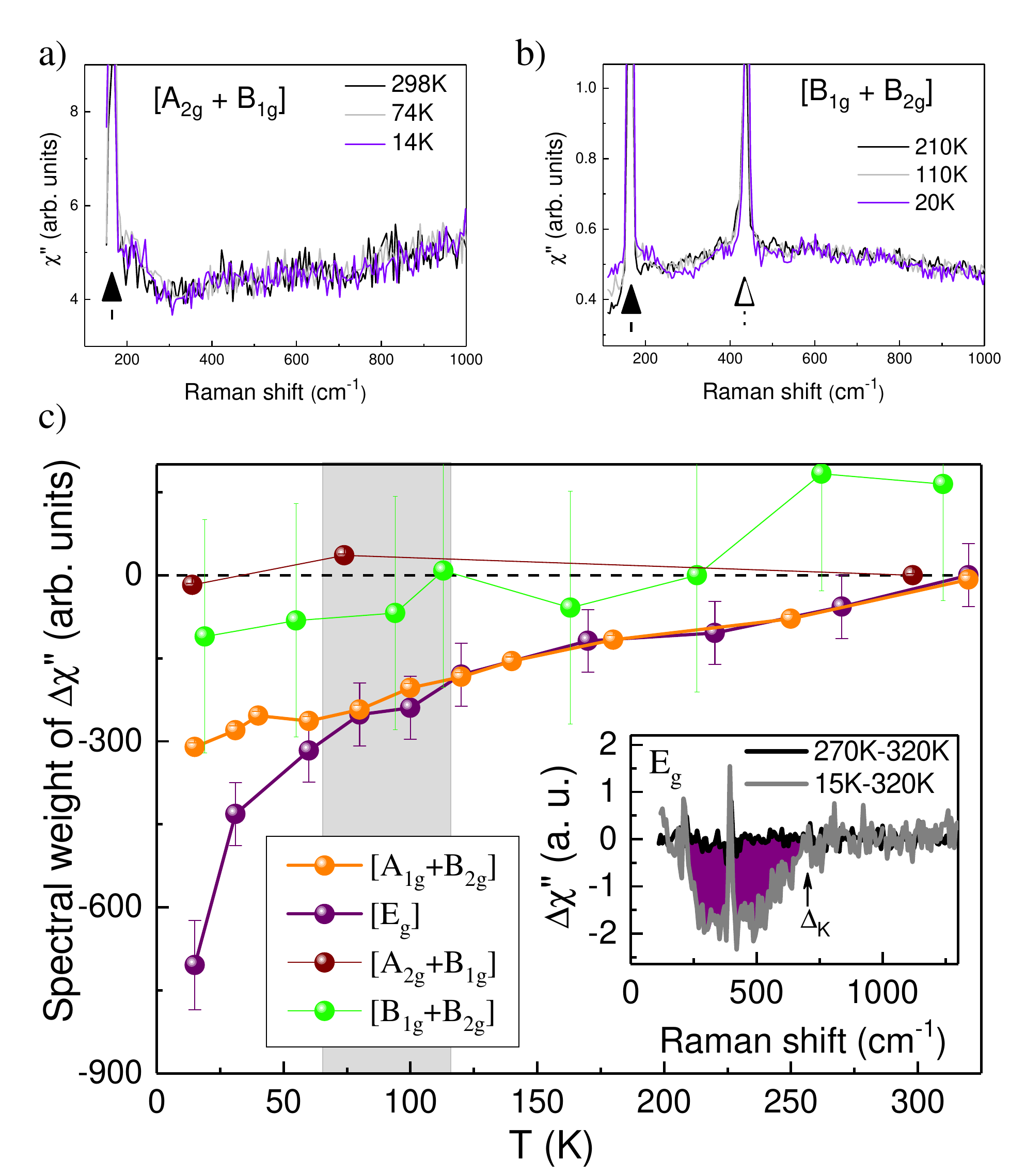}
\caption{(Color online) Raman spectra of \urusi~in the a) [A$_{2g}$ + B$_{1g}$] and b) [B$_{1g}$ + B$_{2g}$] symmetries. The laser line wavelengths are 647~nm and 532~nm, respectively. Open-dotted and full-dashed arrows indicate respectively the A$_{1g}$ (leakage) and the B$_{1g}$ phonon modes. c) Temperature dependence of normalized spectral weights of subtracted Raman responses $\Delta \chi "$=$\chi''(T)-\chi''(T_{ambiante})$ in [A$_{2g}$ + B$_{1g}$], [B$_{1g}$ + B$_{2g}$], [A$_{1g}$ + B$_{2g}$] and [E$_{g}$] symmetries. Insert shows an example of how the spectral weight was extracted in the E$_{g}$ symmetry, in this case at $T=15$~K. A Kondo pseudo-gap opens below $\sim~100$ K only in the E$_g$ symmetry.}
\label{fig2}
\end{figure}

Remarkable effects are observed in the electronic continuum below $\sim$~700\icm~($\sim$ 90 meV) in both the E$_{g}$ and A$_{1g}$ symmetries, i.e. a depletion in a large energy range upon cooling. As shown in Fig.~\ref{fig2} a) and b), such depletion is absent in the B$_{1g}$, B$_{2g}$ or A$_{2g}$ channels. In Figure~\ref{fig2} c), we illustrate this loss of spectral weight by integrating $\Delta\chi'' (T)$=$\chi''(T)-\chi''(320K)$ from the lowest energy (100\icm) to 1500~\icm~for all symmetries. An example of the raw data is presented in the inset of Fig.~\ref{fig2} c). The large error bars in the [B$_{1g}$ + B$_{2g}$] channel prohibit us from drawing any definitive conclusions for the B$_{2g}$ symmetry. Nevertheless, its spectral weight loss is found to be weaker than for the E$_g$ response. This loss of spectral weight happens from 320~K in both E$_{g}$ and A$_{1g}$ symmetries. Below 100 K however, the depletion in the E$_{g}$ symmetry begins to accelerate, while for A$_{1g}$, it  decreases at a constant rate. This temperature scale of 100 K matches with $T_{K}$ \cite{schmidt_imaging_2010, aynajian_visualizing_2010, palstra_superconducting_1985, palstra_anisotropic_1986}.
Thus, the Raman response of URu$_2$Si$_2$ exhibits a clear signature of the Kondo crossover, but only in the E$_{g}$ symmetry, i.e. a partial depletion in the electronic continuum below $\Delta_K\sim~700\icm$ not associated with any symmetry breaking. This symmetry dependent pseudo-gap implies a strong anisotropy in the Kondo physics in URu$_2$Si$_2$. Optical conductivity also shows anisotropic behavior between the \textit{ab}-plane and \textit{c}-axis 
responses \cite{lobo_optical_2015-1, bachar_detailed_2016}. The \textit{ab}-plane optical conductivity is strongly influenced by the formation of coherent Kondo singlets, reminiscent of what happens in the E$_g$ Raman response, while the \textit{c}-axis optical conductivity decreases continuously below 40~meV with no pseudo-gap signatures through the Kondo crossover, mimicking the Raman response seen in the A$_{1g}$ channel.

Finally, it is worth noting that the total spectral weight loss in the E$_{g}$ channel is most significant between 30 K and 15 K when entering the HO phase. Thus, the development of the HO state - for which the A$_{2g}$ Raman signatures are fully developed below 16 K \cite{buhot_symmetry_2014-1} - reinforces the Kondo pseudo-gap opening. This point will be discussed in more detail later on.

While a single energy scale $\Delta_{K}
=700~cm^{-1}$ for the pseudo-gap is identified in both E$_{g}$ and A$_{1g}$ symmetries (see inset fig.\ref{fig2}c), the temperature dependence of the Raman susceptibilities in these symmetries 
reveals two distinct temperature scales: one above 320~K ($T_{K}^{high}$) and a second at T$_K$. High-temperature Kondo ($T_{K}^{high}>T_K$) behavior has been observed previously in transport measurements at $\sim 
370K$ \cite{schoenes_hall-effect_1987}. $T_{K}^{high}$ may result from crystal electric field effects tuning the degeneracy of the \textit{f} orbitals.  Indeed, it has been shown that $T_K$ for a 4-fold degenerate Kondo model is significantly higher than $T_K$ computed for the 2-fold degenerate equivalent model \cite{hewson_kondo_1997,burdin_multiple_2009}. Thus, a similar Kondo description could interpolate smoothly, for instance, from 4-fold degenerate (almost isotropic) impurities below $T_{K}^{high}$ to 2-fold degenerate (anisotropic effective) impurities below $T_{K}$. 
Consequently, the Kondo physics may be observable in several Raman-active symmetries below $T_{K}^{high}$ - here in both E$_g$ and A$_{1g}$ symmetries - and strengthen only in a specific symmetry - E$_{g}$ in the present case - below T$_K$. 

The energy scale for the spectral weight depletion in the E$_g$ and A$_{1g}$ symmetries is quite consistent with the previous observations of a Kondo pseudo-gap in \urusi~at $\sim500$\icm~by optical conductivity measurements \cite{bonn_far-infrared_1988, levallois_hybridization_2011, 
nagel_optical_2012,guo_hybridization_2012,lobo_optical_2015} or at $\sim250$\icm~by STM 
measurements \cite{schmidt_imaging_2010,aynajian_visualizing_2010}. ARPES 
\cite{boariu_momentum-resolved_2013} and quasi-particle scattering spectroscopy 
\cite{park_observation_2012} reported a lower hybridization pseudo-gap at 11~meV ($\sim
$~90\icm). In Kondo models \cite{lacroix_phase_1979,burdin_coherence_2000,burdin_multiple_2009,riseborough_mixed_2016,nozieres_kondo_2005}, the electronic 
dispersion is characterized by a direct pseudo-gap (with zero transferred momentum) of order an effective hybridization $\Delta$ and an indirect pseudo-gap $\Delta^2/W\sim k_BT_K$ 
where $W$ is the electronic bandwidth. Raman spectroscopy probes the direct Kondo pseudo-gap, while, for example, ARPES measures the indirect Kondo pseudo-gap. In the present case, $\Delta\sim\Delta_K=700~\icm$ and if we assume 
$W=1$ eV, we extract an indirect pseudo-gap of 7.5~meV ($\sim$~60\icm) in agreement with the 
ARPES measurements.
From these phenomenological considerations, the relevant temperature scale is found to be of order T$_K$. Interestingly, this description also accounts for the very large ratio between the Kondo direct pseudo-gap energy and the Kondo crossover temperature $\frac{\Delta_K}{k_B \cdot T_K}$. This ratio is of order 10 in URu$_2$Si$_2$, as found in Kondo insulators \cite{degiorgi_charge_2001,freericks_raman_2001}.

In general and in contrast to optical conductivity measurements, electronic Raman scattering is not expected to follow spectral weight sum rules. However, some models show that interaction-dependent sum rules are possible \cite{freericks_optical_2005}.
For instance, in Kondo insulators such as FeSi or SmB$_6$, Raman electronic spectral weight that is suppressed at low frequencies by the charge gap is primarily recovered within an energy range equal to $6\Delta$. This has been discussed theoretically in the framework of the Hubbard model \cite{freericks_raman_2001} and extended to metallic systems for which a clear redistribution of spectral weight to a Fermi-liquid peak at low energy or in some cases to a high energy charge-transfer peak is expected \cite{freericks_nonresonant_2003}.
However, according to our present and previous \cite{buhot_symmetry_2014-1} results, \urusi~does not show any Raman spectral weight redistribution either above or below the depletion energy scale, which seems to indicate that the \urusi~Raman response does not follow any particular sum rule. Thus, our results call for further theoretical investigations in the context of Kondo metallic systems.

The E$_{g}$ symmetry dependence of the Raman electronic continuum presented above implies \textit{k}-space anisotropy in the quasi-particle excitations. Via the Raman vertex \cite{devereaux_inelastic_2007}, polarized Raman scattering has the ability to provide information on the different regions of the Brillouin Zone probed for each selected symmetry. In order to elucidate the \textit{k}-space dependence of the Kondo physics, we have therefore carried out Raman vertex calculations within the effective mass approximation \cite{devereaux_inelastic_2007, valenzuela_optical_2013} \footnote{See Supplemental Material [http://link.aps.org/ supplemental/...], for details about Raman vertex calculations}.

\begin{figure}[h!]
\centering
\includegraphics[width=1\linewidth]{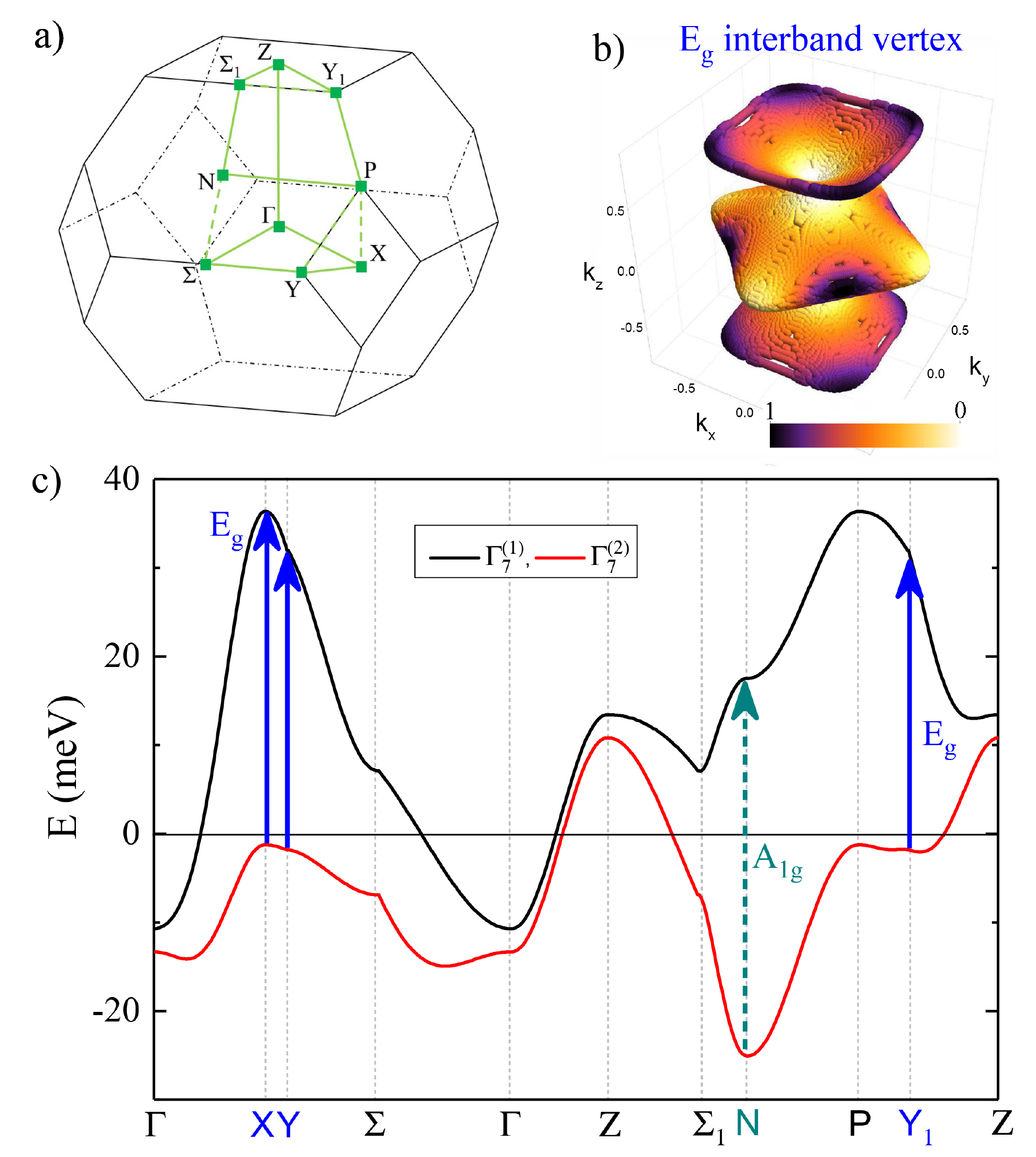}
\caption{(Color online) a) Body-centered tetragonal Brillouin zone of \urusi.
b) Relative magnitude of the E$_{g}$ Raman vertex at the Fermi surface for interband transitions calculated within the effective mass approximation.
c) Calculated electronic dispersion of the $\Gamma^{(1)}_{7}$ and $\Gamma^{(2)}_{7}$ bands including hybridization. 
The solid (dashed) arrows at the X, Y and Y$_{1}$ (N) locations indicate the Brillouin Zone points where the E$_{g}$ ($A_{1g}$) interband vertex reaches a maximum value (See text).}
\label{fig3}
\end{figure}

The intraband and interband Raman vertices have been calculated using a tight-binding electronic band structure developed on the 5\textit{f}$^{2}$ uranium electron's configuration for URu$_{2}$Si$_{2}$ \cite{thalmeier_signatures_2011, rau_hidden_2012}. This model is based on an effective one-particle picture with the \textit{J}=5/2 states. The crystal electric field
effect splits this sextet into three Kramers doublets $\Gamma^{(1)}_{7}\textcircled{+}\Gamma^{(2)}_{7}\textcircled{+}\Gamma_{6}$. For simplicity, the $\Gamma_{6}$ doublet is  neglected. The main contribution to the Fermi surface comes from the two $\Gamma_{7}$ doublets that are described in the tight-binding approximation. Here, the Kondo lattice effect is
considered as a mixing between the two $\Gamma_{7}$ doublets whose contribution to the Hamiltonian depends on a hopping term $t_{12}$ between uranium sites at the zone center and at the corners \footnote{[http://link.aps.org/ supplemental/...], for details about hopping parameters}.
The body-centered tetragonal (BCT) Brillouin zone of the paramagnetic state is shown in Figure \ref{fig3} a).
The Fermi surface obtained (Fig. \ref{fig3} b)) from the electronic dispersion (Fig. \ref{fig3} c)) is consistent with earlier \textit{ab initio} calculations \cite{oppeneer_electronic_2010}. Only the small electrons pockets at $\Gamma$ and $X$  \cite{thalmeier_itinerant_2014} are not reproduced by this simple model \cite{rau_hidden_2012}. The electronic dispersion does not depend strongly on the hopping term $t_{12}$ except around the N point where the energy between the two $\Gamma_{7}$ bands increases when $t_{12}$ deviates from zero \footnote{[http://link.aps.org/ supplemental/...], for details about bands dispersion}.

From our calculations, large values of the Raman vertices were found in the E$_{g}$ and A$_{1g}$ symmetries, while in the B$_{1g}$ and B$_{2g}$ symmetries, the vertices remain very weak \footnote{[http://link.aps.org/ supplemental/...], for details about Raman vertices in all symmetries}. The strongest value of the Raman vertex is in fact found in the E$_{g}$ symmetry for interband transitions with an energy scale of $\sim40$~meV (320$~\icm$), in reasonable agreement with our observations. The E$_{g}$ vertex for intraband transitions, on the other hand, is weak. According to this result, the E$_{g}$ Raman response mainly comes from vertical interband transitions along the diagonals of the Brillouin Zone. Indeed, as emphasized in Figure \ref{fig3} b), the interband E$_{g}$ Raman vertex is strong only at the $X$, $Y$ and $Y_{1}$ points (as indicated by solid arrows in Fig \ref{fig3} c)). Consequently, the Kondo pseudo-gap is linked to the electronic states at these \textit{k}-points, reflecting the \textit{d}-wave like geometry of the Kondo pseudo-gap.

The A$_{1g}$ interband vertex is particularly strong at the N point \footnote{[http://link.aps.org/ supplemental/...], for details about the calculated Raman vertex in the A$_{1g}$ symmetry} where the interband vertical electronic transition has typically the same energy scale. The Raman electronic depletion change is smooth at T$_K$ in this symmetry thus confirming that the Kondo pseudo-gap does not affect the N point of the Brillouin zone; another indication of strong anisotropy in the Kondo physics in URu$_2$Si$_2$.

We now turn to discuss the relationship between the HO state and the Kondo physics in the light of our Raman results. The main observations are two-fold: the Kondo pseudo-gap in the E$_g$ channel is reinforced below T$_{HO}$, confirming that the Kondo effect is not suppressed upon entering the HO state \cite{haule_arrested_2009}.
Moreover, our experimental results reveal no pseudo-gap opening in the A$_{2g}$ symmetry despite the fact that this peculiar symmetry has been shown to exhibit all the Raman signatures of the HO state, including a sharp excitation (at 1.7 meV) and a gap (at 6.8 meV) \cite{buhot_symmetry_2014-1,kung_chirality_2015}.
Raman experiments \cite{cooper_magnetic_1987,buhot_symmetry_2014-1} also showed that the width of the A$_{2g}$ quasi-elastic continuum in the paramagnetic state is affected by the Kondo crossover, suggesting that the A$_{2g}$ Raman channel is sensitive to the hybridization process. 
A definitive conclusion could be reached by going beyond the effective mass approximation so that the A$_{2g}$ Raman vertex is not expected to be zero. This would provide the \textit{k}-dependent variation in the A$_{2g}$ Raman response, notably the HO A$_{2g}$ gap below 6.8\icm, that may also have a \textit{d}-wave geometry and by doing so, may be favorable to the chiral \textit{d}-wave $k_z(k_x+ik_y)$ (E$_{g}$) pairing proposed for the SC state \cite{kasahara_exotic_2007,kawasaki_time-reversal_2014,akbari_hidden-order_2014-1}.

Interestingly, in a two-channel Kondo model such as the hastatic order model \cite{chandra_hastatic_2013}, the normal Kondo
crossover at $T_{K}$ may be converted into a phase transition at $T_{HO}$. Multiple components for the hybridization
gap with different symmetries are expected: 
one that could open at $T_{K}$ (as a crossover without symmetry breaking) and
one that should manifest at $T_{HO}$ as an order parameter for the HO \cite{flint_hidden_2014}. A model inspired by a similar scenario might reconcile the E$_{g}$ symmetry of the Kondo pseudo-gap with the A$_{2g}$ symmetry of the HO gap.

In summary, we report the observation of an anisotropic Kondo pseudo-gap in a metallic Kondo lattice system by Raman spectroscopy. We observe a pseudo-gap opening within the E$_{g}$ symmetry in URu$_2$Si$_2$ that appears to be linked to the Kondo crossover below 100 K. This Kondo pseudo-gap becomes even deeper upon entering the HO state pointing to a link between the Kondo physics and the HO state. From Raman vertex calculations, the interband Raman vertex in the E$_{g}$ symmetry is strongest at the $X$, $Y$ and $Y_1$ points of the Brillouin zone, suggesting that the Kondo pseudo-gap has a \textit{d-wave} like symmetry. This anisotropy in the Kondo physics may play a significant role in the realization of the HO and henceforth should be considered as a key element in further theoretical developments.

\begin{acknowledgments}
This work was supported by the French Agence Nationale de la Recherche
(ANR ''PRINCESS'' Grant No. ANR-11-BS04-0002, ANR-DFG ''Fermi-NEst'' Grant No. ANR-16-CE92-0018, ANR ''SEO-HiggS2'' Grant No. ANR-16-CE30-0014).
We thanks C. Lacroix, J. G. Rau, H.-Y. Kee, P. Thalmeier, L. H. Tjeng, A. Severing, S. Wirth and R. Lobo for very fruitful discussions.
\end{acknowledgments}

\clearpage

\bibliographystyle{apsrev4-1}
\bibliography{biblio}

\end{document}